\documentclass[usenatbib,usegraphicx,onecolumn]{mn2e}
\usepackage{color}

\def\th{\vec{\theta}}

\def\k{{\bf k}}

\def\HI{{\rm HI}}

\def\u{\vec{U}}

\def\k{\vec{k}}

\def\araa{ARAA}

\def\pasp{PASP}
\def\mnras{MNRAS}
\def\prd{Physical Review D,}
\def\na{New Astronomy}

\def\aap{A \& A}

\def\apj{ApJ}

\def\aj{The Astronomical Journal}
\def\u{{\bf U}} 

\def\V2{V_2}
\def\V2ij{V_{2ij}}
\def\S{{\mathcal S}}
\def\V{\mathcal{V}}

\def\la{\left\langle}

\def\lsim{~\rlap{$<$}{\lower 1.0ex\hbox{$\sim$}}}

\def\gsim{~\rlap{$>$}{\lower 1.0ex\hbox{$\sim$}}}

\usepackage{graphics}
\usepackage{color}
\usepackage{amsmath}
\usepackage{amssymb}
\usepackage{psfrag}
\usepackage{subfig}
\usepackage{graphicx}

\newsavebox{\measurebox}
\begin{document}
\date {} 
\title[Image based TGE] {An Image-based Tapered Gridded Estimator (ITGE) for the angular power spectrum} 
\author[S. Choudhuri et al.]{Samir Choudhuri$^{1}$\thanks{Email:samir@ncra.tifr.res.in}, Prasun Dutta$^{2}$ and Somnath Bharadwaj$^{3}$\\
$^{1}$ National Centre For Radio Astrophysics, Post Bag 3, Ganeshkhind, Pune 411007, India\\
$^{2}$ Department of Physics, IIT (BHU), Varanasi 221005, India\\
$^{3}$ Department of Physics,  \& Centre for Theoretical Studies, IIT Kharagpur,  Kharagpur 721 302, India\\}

\maketitle

\begin{abstract}
We present the Image-based Tapered Gridded Estimator (ITGE) to measure
the angular power spectrum $(C_{\ell})$ of the sky signal directly
from the visibilities measured in radio-interferometric
observations. The ITGE allows us to modulate the sky response through
a window function which is implemented in the image plane, and it is
possible to choose a wide variety of window functions. In the context
of the cosmological \HI~21-cm signal, this is useful for masking out
the sky signal from specific directions which have strong residual
foregrounds. In the context of the ISM in external galaxies, this is
useful to separately estimate the $C_{\ell}$ of different parts of the
galaxy. The ITGE deals with gridded data, hence it is computationally
efficient. It also calculates the noise bias internally and exactly
subtracts this out to give an unbiased estimate of $C_{\ell}$. We
validate the ITGE using realistic VLA simulations at $1.4{\rm
  GHz}$. We have applied the ITGE to estimate the $C_{\ell}$ of
\HI~21-cm emission from different regions of the galaxy NGC~628. We
find that the slope of the measured $C_{\ell}$ in the outer region is
significantly different as compared with the inner region.  This
indicates that the statistical properties of ISM turbulence possibly
differ in different regions of the galaxy.
\end{abstract} 


\section{Introduction}
\label{intro}
Measurements of the cosmological redshifted neutral hydrogen (\HI)
21-cm power spectrum can be used to probe the Universe over a large
redshift range $0 < z \lsim 200$
(e.g. \citealt{BA5,furla06,morales10,prichard12,mellema13}). Several
ongoing experiments such as Donald C. Backer Precision Array to Probe
the Epoch of Reionization
(PAPER{\footnote{http://astro.berkeley.edu/dbacker/eor}},
\citealt{parsons10}), the Low Frequency Array
(LOFAR{\footnote{http://www.lofar.org/}}, \citealt{haarlem,yata13}),
the Murchison Wide-field Array
(MWA{\footnote{http://www.mwatelescope.org}},
\citealt{bowman13,tingay13}) and the Giant Metrewave Radio Telescope
(GMRT, \citealt{swarup}) are aiming to detect the 21-cm power spectrum
from the Epoch of Reionization (EoR). Also, future experiment like the
Square Kilometer Array (SKA1
LOW{\footnote{http://www.skatelescope.org/}}, \citealt{koopmans15})
and the Hydrogen Epoch of Reionization Array
(HERA{\footnote{http://reionization.org/}}, \citealt{deboer17})
are designed to detect the EoR 21-cm power spectrum with higher
sensitivity. The major challenges for a detection of the cosmological
21-cm signal from the high redshift Universe are the astrophysical
foregrounds which are 4-5 orders of magnitude brighter than the
expected 21-cm signal \citep{shaver99,santos05,
  ali,paciga11,ghosh1,ghosh2}. Measurements of the foreground power
spectrum are also interesting in their own right. The power spectrum
measurement of the Galactic synchrotron emission can be used to study
the distribution of cosmic ray electrons and the magnetic fields in
the interstellar medium (ISM) of the Milky Way
\citep{Waelkens,Lazarian,iacobelli13}. There are several statistical
detections of the power spectrum of the diffuse Galactic synchrotron
emission at low frequencies which are relevant for the study of the
EoR cosmological 21-cm signal
\citep{bernardi09,ghosh150,iacobelli13,samir17a}.

The power spectrum measurements of other astrophysical signals using
radio interferometric observations are also interesting. The power
spectrum of the continuum emission from supernova remnants (SNRs)
provides information about the statistics of the density and magnetic
field fluctuations in the turbulent plasma of the SNR. \citet{roy09}
have estimated the angular power spectra of the shell-type SNR Cas A
and found that this is consistent with magneto-hydrodynamic turbulence
in the synchrotron emitting plasma. Measurements of the \HI~21-cm
power spectrum from the ISM within our Galaxy and also external
galaxies allow us to probe turbulence on galactic scales.  In a
pioneering study, \citet{crovisier83} have estimated the Galactic
\HI~21-cm angular power spectrum using Westerbork observations and
found a roughly scale-independent power law behaviour. Several
subsequent studies of the \HI~21-cm emission from our Galaxy
\citep{Green93}, the Small Magellanic Cloud and Large Magellanic Cloud
\citep{Stanimirovic99,Kim07} and other external galaxies
\citep{begum06, Dutta08, Dutta09a, Dutta09b, Dutta09c} all find
power-law power spectra, the slopes however vary across these
measurements.  In recent studies \citet{dutta13} and \citet{dutta13na}
have estimated the \HI~21-cm angular power spectrum for a sample of
external galaxies from the THINGS survey and found the power law index
of the power spectrum to vary in the range {\rm -1.9} to {\rm -1.5}
which they interpret as arising from two-dimensional ISM turbulence
spanning length-scale $1$ to $10 \, {\rm kpc}$ in the plane of the
galaxy's disk. Measurements of the \HI~21-cm opacity fluctuations
power spectra \citep{Desh00, Dhawan00, Roy08, Roy12,Dutta14} allow us
to probe the turbulence in the ISM at very small length-scales ($~1 \,
{\rm pc}$ and smaller).

In this endeavour, it is important to choose a suitable estimator to
reliably estimate the power spectrum from the radio-interferometric
data and this is currently an active research area.  \citet{seljak97}
has proposed an image based estimator to measure the polarization in
the cosmic microwave background. The main disadvantage of this image
based estimator is the deconvolution error during image reconstruction
which may affect the estimated power spectrum. \citet{liu12} have
directly used the measured visibilities to estimate the power spectrum
in the context of the cosmological \HI~21-cm signal.  \citet{dillon15}
have introduced a new power spectrum estimation technique to reduce
the error by modelling the covariance of the foreground residuals from
the data itself. The CHIPS estimator developed by \citet{trott16} uses
an inverse-covariance weighting scheme which allows to suppress the
foreground contamination in the measured power spectrum. \citet{liu16}
have developed an estimator which uses the spherical Fourier-Bessel
basis to incorporate the curved sky for large fields of view. The
above-mentioned estimators rely on externally modelling the noise bias
which arises in power spectrum estimation and subtracting this out to
get an unbiased estimate of the power spectrum. \citealt{begum06} and
\citet{Dutta08} have used the two visibility correlation formalism to
estimate the power spectrum and also excluded the self-correlation of
the visibilities which is responsible for the noise bias. Foregrounds
in wide-field radio-interferometric observations pose a severe problem
for detecting the 21-cm power spectrum as it is extremely challenging
to correctly model and subtract bright point sources from the
periphery of the field of view (see \citealt{samir16a} for details).
Residuals point sources from the outer region of the primary beam may
overwhelm the cosmological 21-cm signal \citep{datta10}.
\citet{ghosh1} showed that the point sources located at the periphery
of the main lobe and the side-lobes of the primary beam create an
oscillatory pattern along the frequency direction in the estimated
multi-frequency angular power spectrum.  Equivalently, the wide-field
foregrounds reduce the EoR window by the increasing the area under the
foreground wedge \citep{thyag13}. Using wide-field foreground
simulations, \citet{pober16} showed that it is important to correctly
model and subtract these wide field foregrounds for EoR
detection. \citet{ghosh2} showed that the oscillations in the
multi-frequency angular power spectrum could be suppressed by
restricting the angular extent of the telescope's sky response through
a suitably chosen window function ${\cal W}(\theta)$. This was
implemented by convolving the observed visibilities with
$\tilde{w}(U)$ which is the Fourier transform of ${\cal W}(\theta)$.
\citet{samir14} and \citet{samir16b} have introduced a visibility
based estimator, namely the Tapered Gridded Estimator (TGE) for power
spectrum estimation. TGE uses gridded visibilities to reduce the
computation time and also internally calculates and subtracts the
noise bias to give an unbiased estimate of the power spectrum. The
visibilities are convolved with $\tilde{w}(U)$ at the time of griding
in order to suppress the contribution from the outer regions of the
telescope's field of view.  \cite{samir16a} have used realistic GMRT
simulations to demonstrate that TGE successfully suppresses the
contribution from point sources located at the periphery of the
telescope's field of view and \citet{samir17a} have used the TGE to
estimate the angular power spectrum of the Galactic synchrotron
radiation in two fields of the TIFR GMRT Sky Survey (TGSS)
\citep{Sirothia2014} for which the data was processed and calibrated
by \citet{intema2017}.

The TGE presented in \citet{samir14,samir16b} suppresses the
contribution from the outer region of the telescope's field of view by
convolving the visibilities with $\tilde{w}(U)$ at the time of
gridding. Till now, the various applications of this estimator have
been restricted to situations where the sky response is tapered using
a circularly symmetric Gaussian window function ${\cal
  W}(\theta)=exp(-\theta^2/\theta^2_w)$ where the values of $\theta_w$
have been chosen so as to suppress the sky response towards the
periphery of the main lobe of the primary beam pattern (Figure 1 of
\citealt{samir16a}).  In such cases, $\tilde{w}(U)$ also is a
circularly symmetric Gaussian and the extent of the convolution is
restricted to a small disk in the baseline plane, the value of
$\tilde{w}(U)$ becomes extremely small beyond this disk, and the
contributions from the distant baselines can be neglected. The TGE
proves to be a fast and reliable power spectrum estimator in such
situations.  However, there are situations where one would like to use
a window function ${\cal W}(\th)$ which is not a simple circularly
symmetric Gaussian. In the context of the cosmological \HI~21-cm power
spectrum, it would be preferable to use a window function which has a
value ${\cal W}(\theta) \sim 1$ over a reasonably large angular extent
near the centre of the field of view and then vary rapidly fall to a
value close to zero at larger angles (for example the Butterworth
function used here) instead of the Gaussian which falls off gradually
away from the centre. Again, it may be desirable to mask out select
regions of the sky corresponding to the locations of bright sources
where significant residuals persist after source subtraction
(e.g. Figure 9 of \citealt{ghosh150}). In the context of the ISM, we
may be interested in separately measuring the power spectrum in
different parts of the galaxy. For example, the power spectrum in the
star-forming region in the inner parts of the galaxy may be different
as compared to that in the outer parts. In all of these cases ${\cal
  W}(\th)$ ceases to be a circularly symmetric Gaussian function, and
quite often we do not even have a closed-form analytic expression for
$\tilde{w}(U)$.  Further, the function $\tilde{w}(U)$ then covers a
large extent in the baseline plane and the convolution of the
visibilities is computationally expensive. In such situations, it is
advantageous to directly apply the window function ${\cal W}(\th)$ in
the image plane instead of convolving the visibilities. The noise in
the different image points are however correlated. The problem "How to
avoid the noise bias?" however still persists. In this paper, we have
formulated an image-based Tapered Gridded Estimator (ITGE) where we
replace the convolution in the visibility plane by a multiplication in
the image plane. We have used the
  FFTW{\footnote{http://www.fftw.org}} to make image making
computationally fast. ITGE also calculates the noise bias internally
and subtracts this out to give an unbiased estimate of the power
spectrum. In this paper, we apply the ITGE to \HI~21-cm data of the
galaxy NGC~628 to separately estimate the angular power spectrum of
the ISM in the inner and outer parts of this galaxy.

A brief outline of the paper follows. In Section 2, we present the
mathematical formalism for ITGE. In Section 3, we validate this
estimator using realistic simulations. In Section 4, we present an
application of this estimator to determine the angular power spectrum
of the galaxy NGC~628. Finally, we summarize and conclude in Section
5.

\section{Mathematical formulation of the Image Based Tapered Gridded Estimator}
\label{form}
In this section, we present the mathematical formalism leading to the
ITGE. Here, we adopt a notation to represent the transformation
between the visibilities in the baseline plane and the image on the
sky plane. We deal with gridded data throughout, and these
transformations are discrete Fourier transforms which can be
implemented using either DFT or FFT. In our notation, the value of a
quantity $L$ at the grid point $a$ on the image plane is denoted as
$L_{a}$, and the corresponding quantity $Q$ at the grid point $g$ on
the baseline plane is denoted as $Q_{g}$. We denote the forward and
backward Fourier transform between $L$ and $Q$ as $Q_{g} = \mathcal{F}
[L_{a}]$ and $L_{a} = \mathcal{I} [Q_{g}]$ respectively. Further, here
we restrict our discussion to single frequency observations for
simplicity.

The basic idea of ITGE is the same as that of the visibility based TGE
where the estimator ${\hat E}_g$ at the grid point $g$ in the baseline
plane is defined through
\begin{equation}
{\hat E}_g = M_g^{-1} \, \left ( \mid \V_{cg} \mid^2 - \sum_i \mid
\tilde{w} (\u_g-\u_i) \mid^{2} \mid \V_{i} \mid^{2}\right ) \,.
\label{eq:a1}
\end{equation}
as given in eqn.~(17) of \citet{samir16b}. The expectation value of
the estimator with respect to different random realizations of the
visibility signal gives an unbiased estimate of the angular power
spectrum $C_{\ell_g}$ at the angular multipole $\ell_g = 2 \pi U_g$
corresponding to the baseline $\u_g$.  Here $M_g$ is a normalization
factor, $\V_{i}$ is the visibility measured at the baseline $\u_i$ and
$\V_{cg}$ represents the convolved visibilities which is evaluated at
the grid point $g$ using
\begin{equation}
\V_{cg} = \sum_{i}\tilde{w}(\u_g-\u_i) \, \V_i \,.
\label{eq:a2}
\end{equation}

As mentioned earlier, this convolution effectively multiplies the
telescope's sky response with the window function ${\cal W}(\th)$
resulting in a tapered field of view. It is possible to directly
estimate the power spectrum using $M_g^{-1} \, \left ( \mid \V_{cg}
\mid^2 \right)$, however the self-correlation of the individual
visibilities introduces a positive noise bias in the estimated power
spectrum. The second term in eqn.~(\ref{eq:a1}) - $\sum_i \mid
\tilde{w} (\u_g-\u_i) \mid^{2} \mid \V_{i} \mid^{2}$ evaluates the
contribution from this self-correlation and subtracts this out to give
an unbiased estimate of the angular power spectrum. Some signal also
is lost, however, this is expected to be small when the number of
visibilities at each grid point is large.  The value of the
normalization factor $M_g, $ at each grid point, depends on the actual
baseline distribution, the antenna primary beam pattern and the
tapering window function. As discussed in \citet{samir16b}, we
simulate the visibilities for a unit angular power spectrum (UAPS)
$(C_{\ell}=1)$ sky signal and use these to estimate the values of
$M_g$.

We now discuss how we have implemented eqn.~(\ref{eq:a1}) in the image
domain.  We proceed by first gridding the visibilities using
\begin{equation}
\V_{g} = \sum_i \Theta \left ( 1 - \frac{\mid \u_{g} - \u_{i}
  \mid}{(\Delta U/2)} \right ) \V_{i} c_{i}.
\label{eq:a3}
\end{equation}
Here $\Theta (\u)$ is the Heaviside step function, $\Delta \u$ denotes
the baseline grid spacing which has been chosen to be sufficiently
small so that the angular scale $(\Delta U)^{-1}$, which is the extent
of the dirty image, is much larger than the telescope's field of
view. In this paper, we choose $\Delta U=0.2D/\lambda$ which
corresponds to an image of dimension $5$ times the telescope's field
of view. Here $c_{i}$ is a weighting function whose value we have the
freedom to choose. There are two common choices of the weighting
function on a grid: (a) the natural weighting which gives a constant
weight $c_{i}=1$ to all the visibilities, and (b) the uniform
weighting which gives a weight $c_{i}=1/N_g$ which is inversely
proportional to the sampling density {\it i.e. } the number of
visibilities $N_g$ at a particular grid point $g$. The choice of
weighting scheme introduces differences in the resulting image, and
details of possible weighting schemes are discussed in
\citet{thompson}. Here we have used $c_{i}$=1 {\it i.e.}  the natural
weighting throughout the paper. We use the gridded visibilities to
calculate the dirty image $I_{a}$ using
\begin{equation}
I_{a} = \mathcal{I} [\V_{g}],
\label{eq:a4}
\end{equation}
and determine the convolved visibilities $\V_{cg}$ using
\begin{equation}
\V_{cg} = \mathcal{F} [{\cal W}_{a} I_{a} ] \,.
\label{eq:a5}
\end{equation}
Here ${\cal W}(\th)$ is the window function that we wish to implement
on the sky, and it is not restricted to be a circularly symmetric
Gaussian function.

The issue now is to estimate the contribution from the visibility
self-correlation using operations in the image plane.  We proceed by
first gridding the visibility self-correlation using
\begin{equation}
VSQ_{g} = \sum_i \Theta \left ( 1 - \frac{\mid \u_{g} -
  \u_{i}\mid}{(\Delta \u/2)} \right ) \mid \V_{i} \mid^{2} c_{i}^{2}.
\label{eq:a6}
\end{equation}
We now consider $\tilde{w}_g={\cal F}[{\cal W}_a]$ which is the
Fourier transform of the window function ${\cal W}(\th)$, and
construct the images of $ \mid \tilde{w}(\u) \mid^2$ and the
visibility self-correlation (eqn.~\ref{eq:a6}) using
\begin{equation}
WSQ_{a} = \mathcal{I} \left [ \mid \tilde{w}_g \mid^2 \right]
\label{eq:a7}
\end{equation}
and
\begin{equation}
ISQ_{a} = \mathcal{I} \left [ VSQ{g} \right ]
\label{eq:a7a}
\end{equation}
respectively. It is possible to obtain the second term in
eqn~(\ref{eq:a1}) by taking the Fourier transform of the product of
the two images $WSQ$ and $ISQ$ {\it i.e.}
\begin{equation}
\sum \limits _i \mid \tilde{w} (\u_{g} - \u_{i} ) \mid^{2} \mid \V_{i}
\mid^{2} = \mathcal{F} \left[ ISQ_{a} \, WSQ_{a} \right].
\label{eq:a8}
\end{equation}
We have used eqns.~(\ref{eq:a5}) and (\ref{eq:a8}) in
eqn.~(\ref{eq:a1}) to define the ITGE. As for the TGE, we have used
UAPS simulations to estimate the normalization factor $M_g$. The
values of the angular power spectrum estimated at different grid
points are binned to increase the signal to noise ratio and for
convenience of displaying and interpretation. Considering annular bin
labeled using $a$, we have the binned ITGE defined as
\begin{equation}
\hat{E}(a) = \frac{\sum\limits_g a_{g} \hat{E}_{g}}{\sum\limits_g
  a_{g}}
\label{eq:a9}
\end{equation}
where $a_{g}$ is the weight assigned to any particular grid
point. This provides an estimate of the bin averaged angular power
spectrum $\langle \hat{E}(a) \rangle = C_{\ell_a}$ at the mean angular
multipole
\begin{equation}
\ell_a=\frac{\sum\limits_g a_g \ell_g}{\sum\limits_g a_g} \,.
\end{equation}
Here we have evaluated the angular power spectrum in equally spaced
logarithmic bins, and assigned equal weights to all the grid points
i.e, $a_{g}=1$.

\section{Validation}
\label{valid}
We first describe the simulation of radio interferometric visibilities
which we have used to validate the ITGE. To simulate the sky signal we
follow the same procedure as described in
\citep{samir14,samir17b}. Here, we use a model angular power spectrum
\begin{equation}
C^M_{\ell}=A \times \left(\frac{1000}{\ell} \right)^{\beta},
\label{eq:b1}
\end{equation}
where $A=3\times10^5$ and we choose the value of the power-law index
$\beta=1.88$, this choice of values is guided by the measured
$C_{\ell}$ for the galaxy NGC~628 which we analyze in the next
Section. We generate the Fourier components of the temperature
fluctuations on a two-dimensional grid using this model angular power
spectrum and then use the FFTW to generate the temperature
fluctuations $\delta T(\th)$ on the sky plane.  The total number of
grid points used in this simulation are $1024\times1024$ with an
angular resolution $0.045^{'}$. The flat-sky approximation is
  valid for a region of size $\sim 46^{'}$ considered here. We use
the baseline configuration of VLA antennas at $1.4 \ {\rm GHz}$
pointed towards the direction of the galaxy NGC~628 (R.A.=$01{\rm h}
\, 36{\rm m} \, 41{\rm s}$ Dec=$15^{\circ} \, 47^{'} \, 00^{''}$) to
simulate the visibilities. We first multiply the simulated $\delta
T(\th)$ with the VLA primary beam pattern and then use 2-D FFTW to
calculate the visibilities on a baseline grid. These gridded values
were interpolated to calculate the visibilities along the simulated
baseline $uv$ tracks. Details of similar simulations have been
discussed in \citep{samir17b}.

\begin{figure}
\begin{center}
\includegraphics[width=75mm,angle=0]{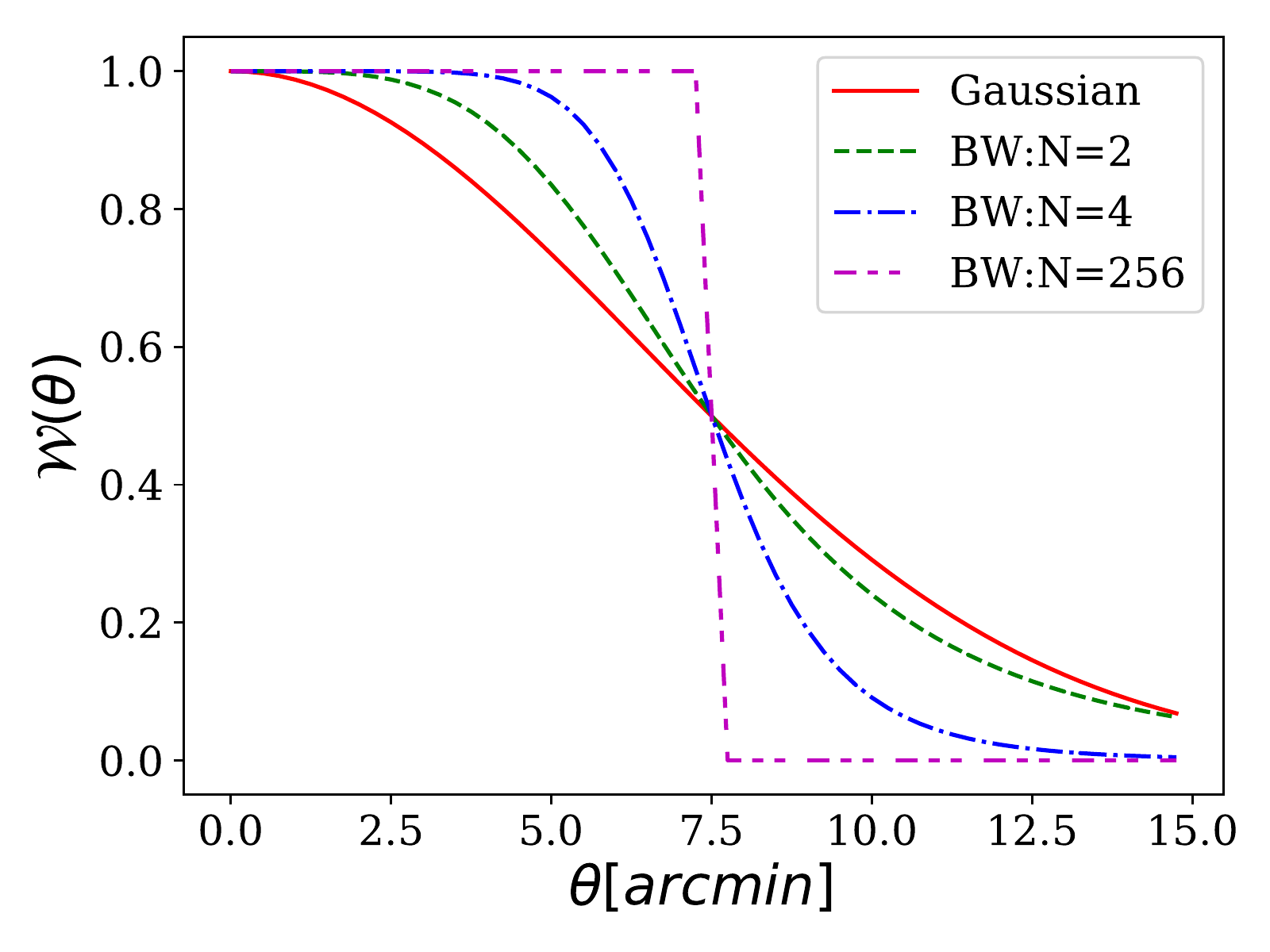}
\caption{This shows the different window functions (Gaussian and
  Butterworth (BW)) used in our analysis.  All of the window function
  have a HWHM of $7.5^{'}$}
\label{fig:fig1}
\end{center}
\end{figure}

To validate the estimator ITGE, we choose different types of window
functions as shown in Figure \ref {fig:fig1}.  We first consider a
Gaussian window
\begin{equation}
{\cal W}_G(\th)=\exp[-\theta^2/\theta^2_g]
\label{eq:b3} 
\end{equation}
with $\theta_g=9^{'}$ as shown by the red solid curve in Figure
\ref{fig:fig1}.  The half width half maxima (HWHM) of this Gaussian
window is $\theta_{HWHM}=\sqrt{\ln2} \, \theta_g=7.5^{'}$.  We also
consider the Butterworth (BW) window given by
\begin{equation}
{\cal W}_{BW}(\th)=\frac{1}{1+({\theta}/{\theta_b})^{2N}}
\label{eq:b4} 
\end{equation}
Here $\theta_b$ is the HWHM of the BW window function, and we choose
this to be $\theta_b=7.5^{'}$ which is the same as that of the
Gaussian. The index $N$ determines how sharply the window falls as a
function of the angular distance $\theta$. Figure \ref {fig:fig1}
shows the shape of the BW window for different values of $N~(=2,4$ and
$256)$. We see that compared to the Gaussian, the BW window is flatter
in the central region and falls off more sharply at large angles
beyond the HWHM. Further, the BW window function falls more sharply as
the value of $N$ is increased.

\begin{figure}
\begin{center}
\includegraphics[width=85mm,height=70mm,angle=0]{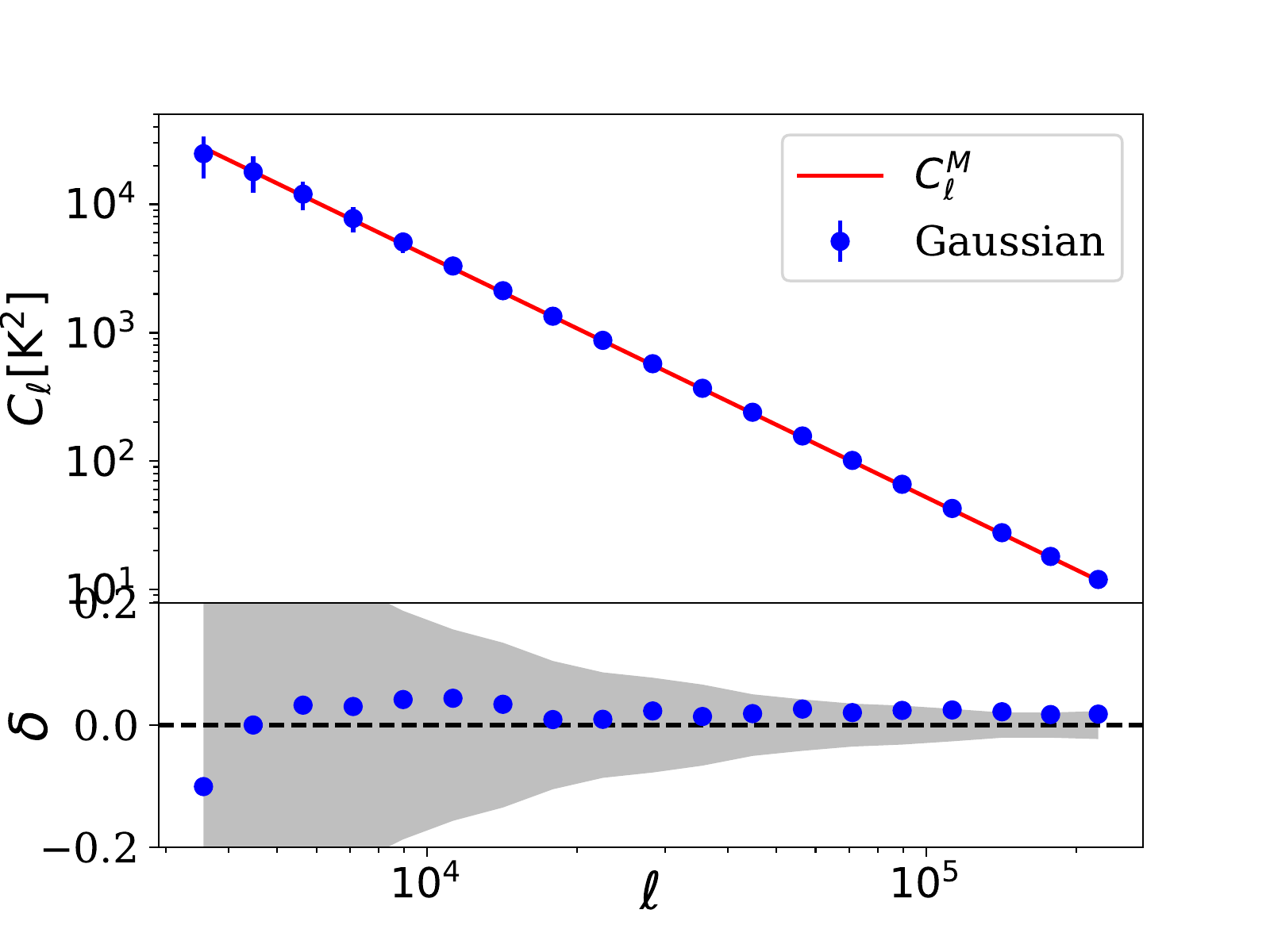}
\includegraphics[width=85mm,height=70mm,angle=0]{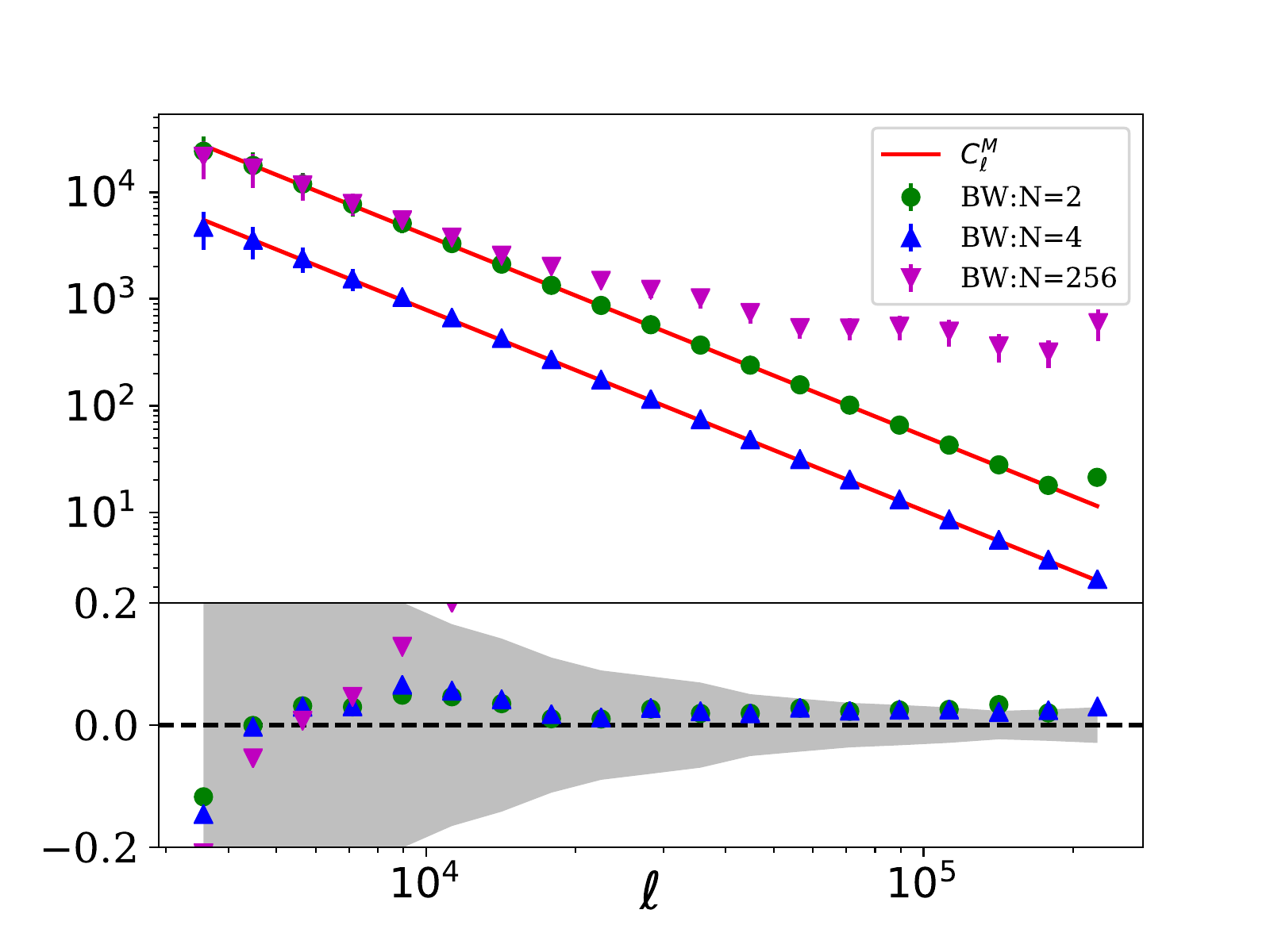}
\caption{The left and right panels show the results for the Gaussian
  and the Butterworth tapering windows respectively. The upper panels
  show the estimated $C_{\ell}$ with $1-\sigma$ error bars. We also
  show the model $C^M_{\ell}$ with red solid line for comparison. The
  lower panels show the fractional deviation $(\delta)$ between the
  estimated $C_{\ell}$ and model $C^M_{\ell}$. The shaded region in
  the lower panels show the expected statistical fluctuations
  $\sigma/C^M_{\ell}$. In the right panel we use different color for
  different BW windows: $N=2$ (green circles), 4 (blue upper
  triangles) and 256 (magenda lower triangles). For $N=4$, we have
  scales the values of the estimated $C_{\ell}$ and also the model
  $C^M_{\ell}$ for clarity of presentation in the right upper panel.}
\label{fig:fig2}
\end{center}
\end{figure}

We apply the ITGE to the simulated visibilities to measure the angular
power spectrum considering all the window functions shown in Figure
\ref {fig:fig1}. Figure \ref {fig:fig2} shows the mean estimated
$C_{\ell}$ and the rms. fluctuation $\sigma$ calculated using
  $1000$ independent realizations of the simulated data. The estimated
  $\sigma$ will increase if we reduce the width of the window
  function. This is due to the fact that the smaller window reduces
  the sky response, effectively causing large cosmic variance. In
  comparison, the HWHM of all the windows shown in Figure \ref
  {fig:fig1} are the same, and we expect the cosmic variance to be
  quite similar for these windows. Here we use $20$ equally spaced
logarithmic bins in the $\ell$ range from $3\times10^3$ to
  $3\times10^5$ to increase the signal to noise ratio in the
estimated $C_{\ell}$ values. The left panels show the results for the
Gaussian tapering window. The upper left panel shows the estimated
$C_{\ell}$ with $1-\sigma$ error bars (blue solid circles). We also
show the input model $C^M_{\ell}$ (eqn. \ref{eq:b1}) with a red solid
line for comparison.  We see that the estimated values are in
reasonably good agreement with the model for $\ell \ge
  3\times10^3$, the tapering due to the primary beam pattern and the
window function modifies the shape of the estimated power spectrum
at $\ell < 3\times10^3$ \citep{samir14}, and we have not shown
  this range in these figures. We also fit the estimated $C_{\ell}$
  with a power law (eq. \ref{eq:b1}) to see the efficacy of the
  estimator and the best fit value of the parameter $\beta$ is
  $1.87\pm0.002$ which is close to the input model $(\beta=1.88)$ used
  for the simulations .  Table ~\ref{tab:1} summarizes the recovered
  $\beta$ for all the window functions used in this work. The lower
left panel shows $\delta=(C_{\ell}-C^M_{\ell})/C^M_{\ell}$ which is
the fractional deviation between the estimated $C_{\ell}$ and the
input model $C^M_{\ell}$. We see that the fractional deviation is
  less than $10 \%$ for all $\ell$ values in the range
  $\ell\ge3\times10^3$.  The shaded region in the lower panel shows
  the expected statistical fluctuations $\sigma/C^M_{\ell}$. We see
  that the fractional deviation $\delta$ is within $1-\sigma$
  statistical fluctuations for the whole $\ell$ range $3\times10^3 \le
  \ell \le 3\times10^5$ considered here.
\begin{table*}
\begin{center}
\begin{tabular}{c c c c c c c c}
\hline &Model&Gaussian&BW,N=2 &BW,N=4 &BW,N=256&Annulus& Mask
\\ \hline $\beta$ & $1.88$ & $1.87\pm0.002$ & $1.87\pm0.002$ &
$1.87\pm0.003$ & Not recovered & $1.87\pm0.007$ & $1.87\pm0.01$
\\ \hline
\end{tabular}
\caption{This shows the recovered $\beta$ for the different
    window functions used in this work. Here we have used $\beta=1.88$
    as the input model to simulate the sky. For the BW window with
    $N=256$ the estimated $C_{\ell}$ shows considerable deviations
    from a power law.}
\label{tab:1}
\end{center}
\end{table*}

The right panels of Figure \ref {fig:fig2} show the results for the BW
windows. Here we use $N=2, 4$ and $256$ for which the shape of the
window functions are shown in Figure \ref {fig:fig1}. The upper right
panel shows the estimated $C_{\ell}$ with $1-\sigma$ error bar for
$N=2$ (green circles), $4$ (blue upper triangles) and $256$ (magenta
lower triangles). We also show $C^M_{\ell}$ with a red solid line for
comparison. For $N=4$, we have scaled the values of the estimated
$C_{\ell}$ and also $C^M_{\ell}$ for clarity of presentation.  We see
that for $N=2$ and $4$ the estimated $C_{\ell}$ matches quite well
with the model for the whole $\ell$ range. For $N=256$, the estimated
$C_{\ell}$ deviates significantly from $C^M_{\ell}$ at large $\ell$
values, and we are able to recover $C^M_{\ell}$ within a limited
  $\ell$ range $3\times10^3\le\ell\le2\times10^4$. In this case
($N=256$) the window function falls sharply at $\theta=7.5^{'}$ (the
HWHM, Figure \ref{fig:fig1}) which introduces oscillations in the
Fourier domain. The large deviations in the estimated $C_{\ell}$ are
possibly a consequence of these oscillations, and we subsequently
restrict the value of $N$ to $2$ and $4$ where the window function
does not fall so sharply. We also fit the estimated $C_{\ell}$
  for $N=2$ and $4$ with a power law and the best fit values of
  $\beta$ are shown in Table ~\ref{tab:1}.  The lower right panel
shows the fractional deviation $\delta$ for the BW windows.  Like for
the Gaussian window, the deviations $\delta$ are consistent with the
$1-\sigma$ statistical fluctuations for BW windows with $N=2$ and $4$.
We note that the values of $\delta$ are less than $10 \, \%$ for the
whole range considered here.

\begin{figure}
\begin{center}
\includegraphics[width=85mm,height=70mm,angle=0]{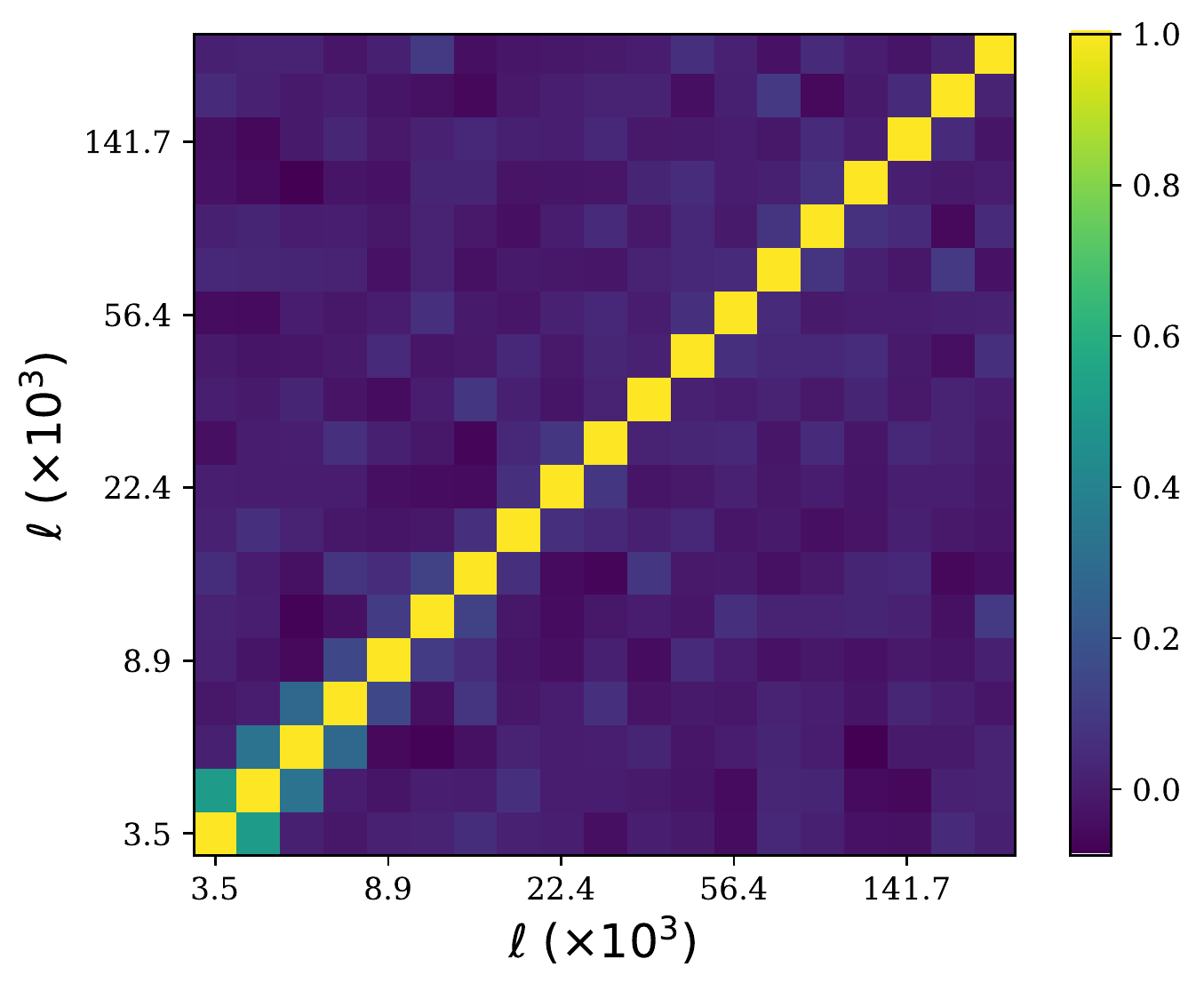}
\includegraphics[width=85mm,height=70mm,angle=0]{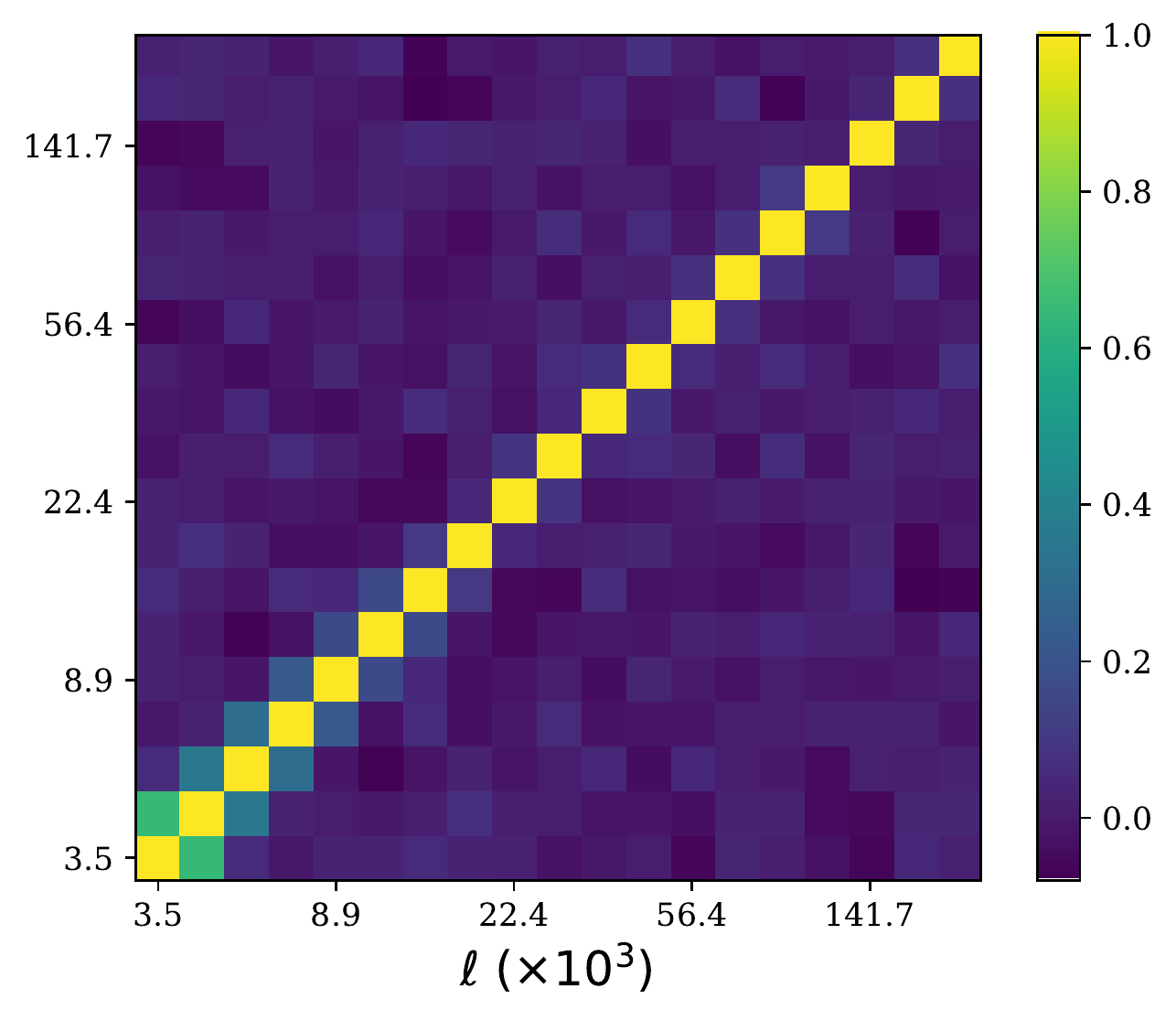}
\caption{The left and right panels show the correlation
    coefficient, $r_{ij}$ between different $\ell$-bins for the
    Gaussian and the BW (N=4) window functions respectively.  The
    correlation properties for both these windows are quite similar.
    At low $\ell$ $(\le 9 \times 10^3)$ the two adjacent bins show
    correlation $r_{ij}\sim 0.5$. The correlation is small ($\mid
    r_{ij} \mid \sim 0.1$ or smaller) elsewhere.}
\label{fig:fig2a}
\end{center}
\end{figure}

The ITGE allows us to select particular regions of the image
  during $C_{\ell}$ estimation by choosing a suitable window
  function. The extent of the window function $\tilde{w}(\u)$ in
  Fourier space introduces a correlation between the power spectrum
  estimates at different $\ell$ bins. This could be particularly
  important when the bin spacing is smaller than the extent of the
  window function. To quantify this we have calculated the error
  covariance $C_{ij}=\langle \delta P_i \, \delta P_j \rangle$ of the
  power spectrum in the $i$ and $j$-th bins. We use the correlation
  coefficient $r_{ij}=C_{ij}/(\sigma_i \sigma_j)$ where $\sigma^2_i$
  and $\sigma^2_j$ are the error variance of the estimated power
  spectrum in the $i$ and $j$-th bins respectively.  As mentioned
  earlier, the variance and covariance were estimated using $1,000$
  independent realizations of the simulations.  Figure \ref
  {fig:fig2a} shows $r_{ij}$ for the Gaussian window function (left
  panel) and the BW window function (right panel) with $N=4$ which we
  have used in the subsequent analysis. We see that in both cases
  there is some correlation $(r_{ij}\sim 0.5)$ between the two
  adjacent bins (one on each side) at small $\ell$ $(
  \le9\times10^3)$, whereas the $\ell$-bins are all uncorrelated at
  larger $\ell$ where $\mid r_{ij} \mid$ has values $\sim 0.1$ or
  smaller. This is consistent with the fact that the window has a
  width of $~7.5^{'}$ in the image plane. This has an extent in
  Fourier space which corresponds to $\Delta \ell \simeq 3,000$.

\begin{figure}
\begin{center}
\includegraphics[width=50mm,angle=0]{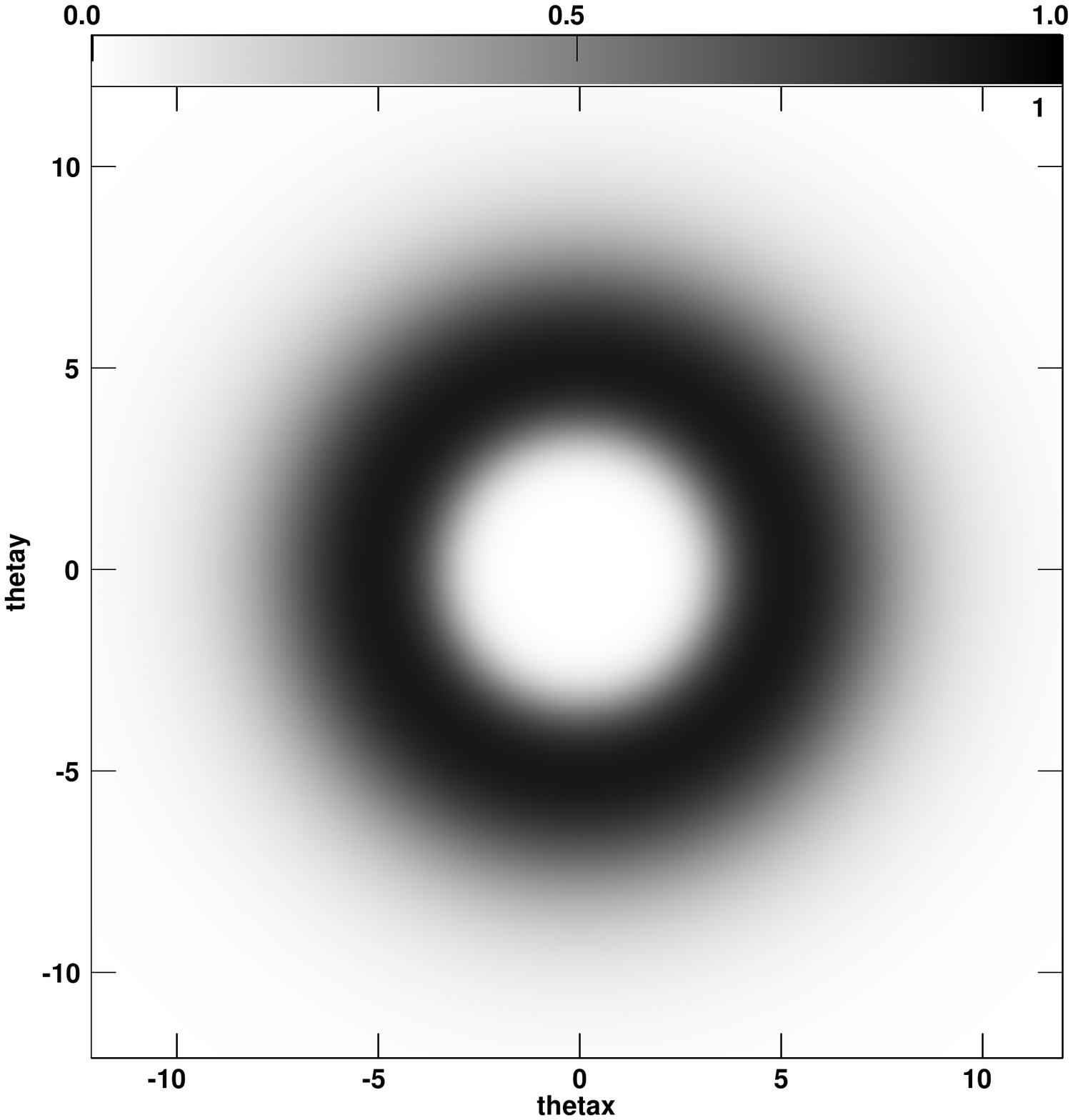}
\includegraphics[width=50mm,angle=0]{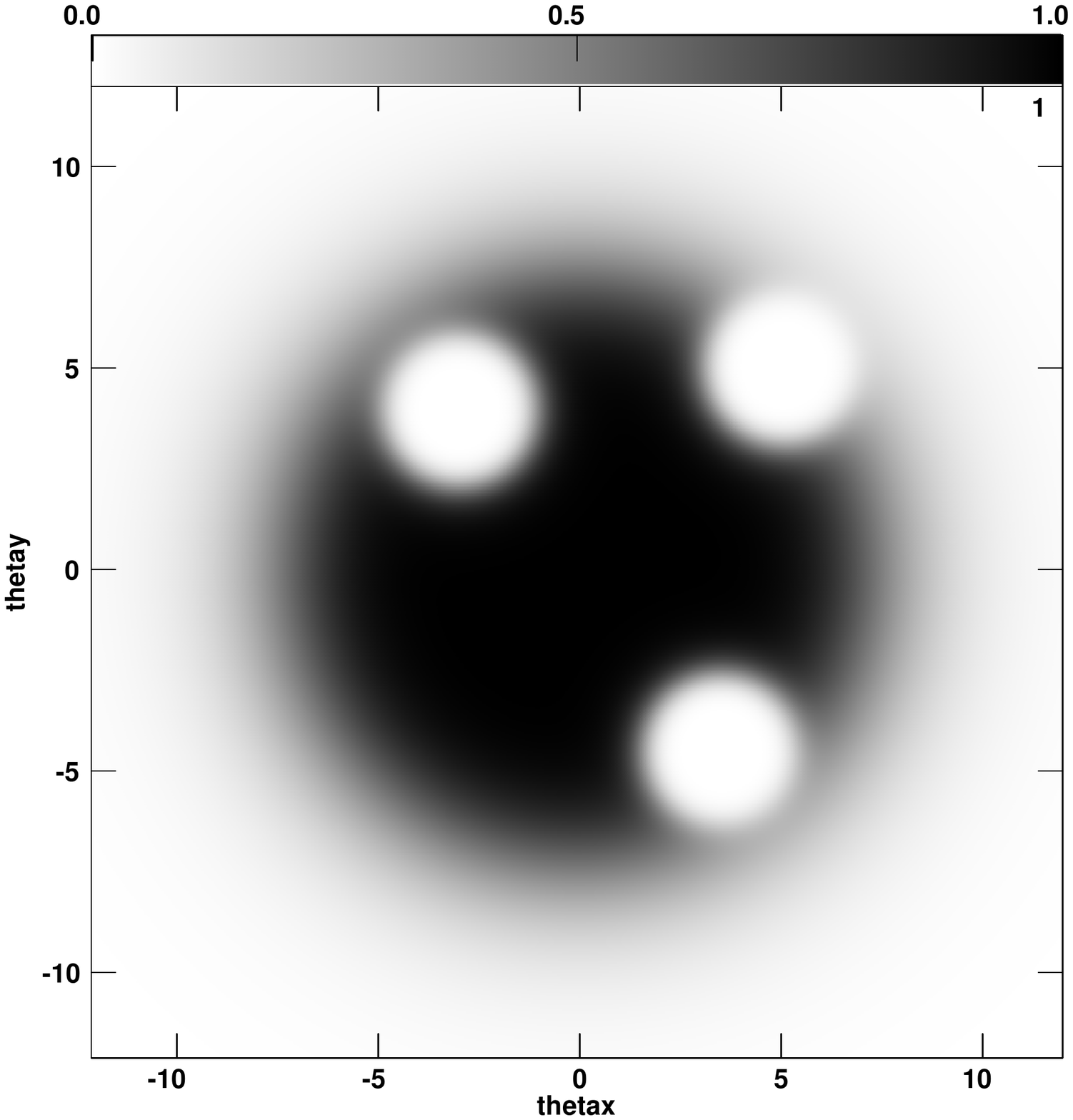}
\includegraphics[width=75mm,height=65mm,angle=0]{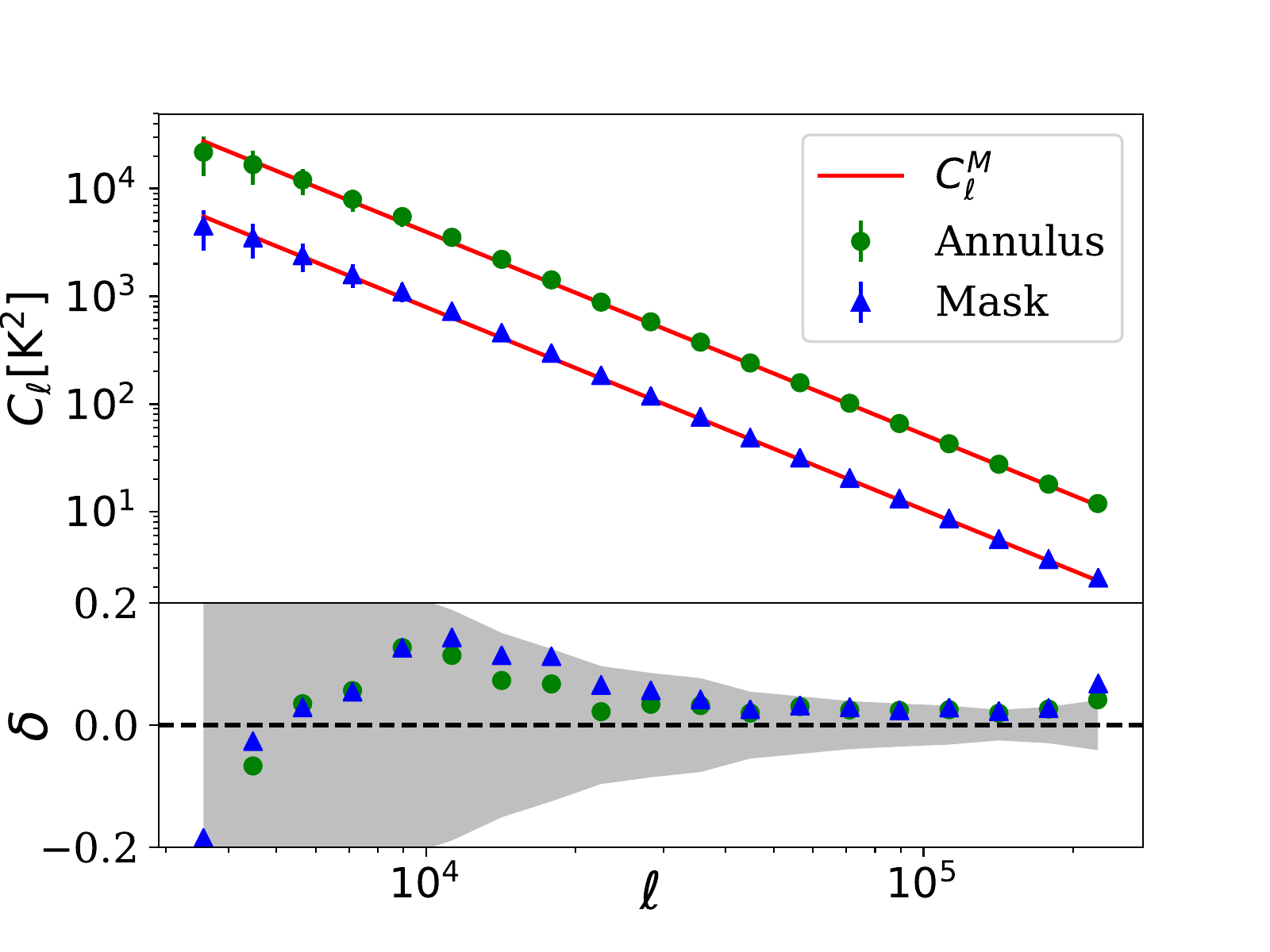}
\caption{The left panel shows the Annulus window with inner HWHM
  radius $3.5^{'}$ and outer HWHM radius $7.5^{'}$. The middle panel
  shows the Mask window where we place three masks with angular HWHM
  radius $2^{'}$ at different locations on top of the BW window which
  has $\theta_b=7.5^{'}$. In both cases we use $N=4$ for the BW
  window. The right panel shows the same as Figure \ref {fig:fig2} but
  for the Annulus and Mask windows.}
\label{fig:fig3}
\end{center}
\end{figure}

We have also validated the estimator for two other window functions
namely (a) the Annulus window, and (b) the Mask window. The Annulus
window, shown in the left panel of Figure \ref {fig:fig3}, allows us
to estimate $C_{\ell}$ only from the outer annular region. The Annulus
window considered here is a combination of two $N=4$ BW windows, an
inner window with HWHM radius $3.5^{'}$ and outer window with HWHM
$7.5^{'}$, and only the annular region between the two windows is used
for the analysis. The Mask window, shown in the middle panel of the
figure, has a $N=4$ BW window with $\theta_b=7.5^{'}$.  In addition to
this, three $N=4$ BW windows with $\theta_b=2^{'}$ have been used to
mask the radiation from three specific directions. The upper right
panel of the figure shows the estimated $C_{\ell}$ along with the
input model $C^M_{\ell}$. The results have been arbitrarily scaled for
clarity of presentation. The lower right panel compares the fractional
deviation $\delta$ with the statistical fluctuations shown by the
shaded region.  We see that for both the window functions the
  estimated $C_{\ell}$ is largely in agreement with the model
  $C^M_{\ell}$ for the whole $\ell$ range. Again, we fit the estimated
  $C_{\ell}$ with a power law and the best fit values of $\beta$ are
  $1.87\pm0.007$ and $1.87\pm0.01$ for the Annulus and the Mask
  windows respectively (Table ~\ref{tab:1}). The fractional deviation
  is less than $15 \, \%$ for these windows. The fractional deviation
  $\delta$ is also consistent with the statistical fluctuations
  $(\sigma/C^M_{\ell})$ which is shown by the shaded region in the
  right lower panel.

\begin{figure}
\begin{center}
\includegraphics[width=85mm,height=70mm,angle=0]{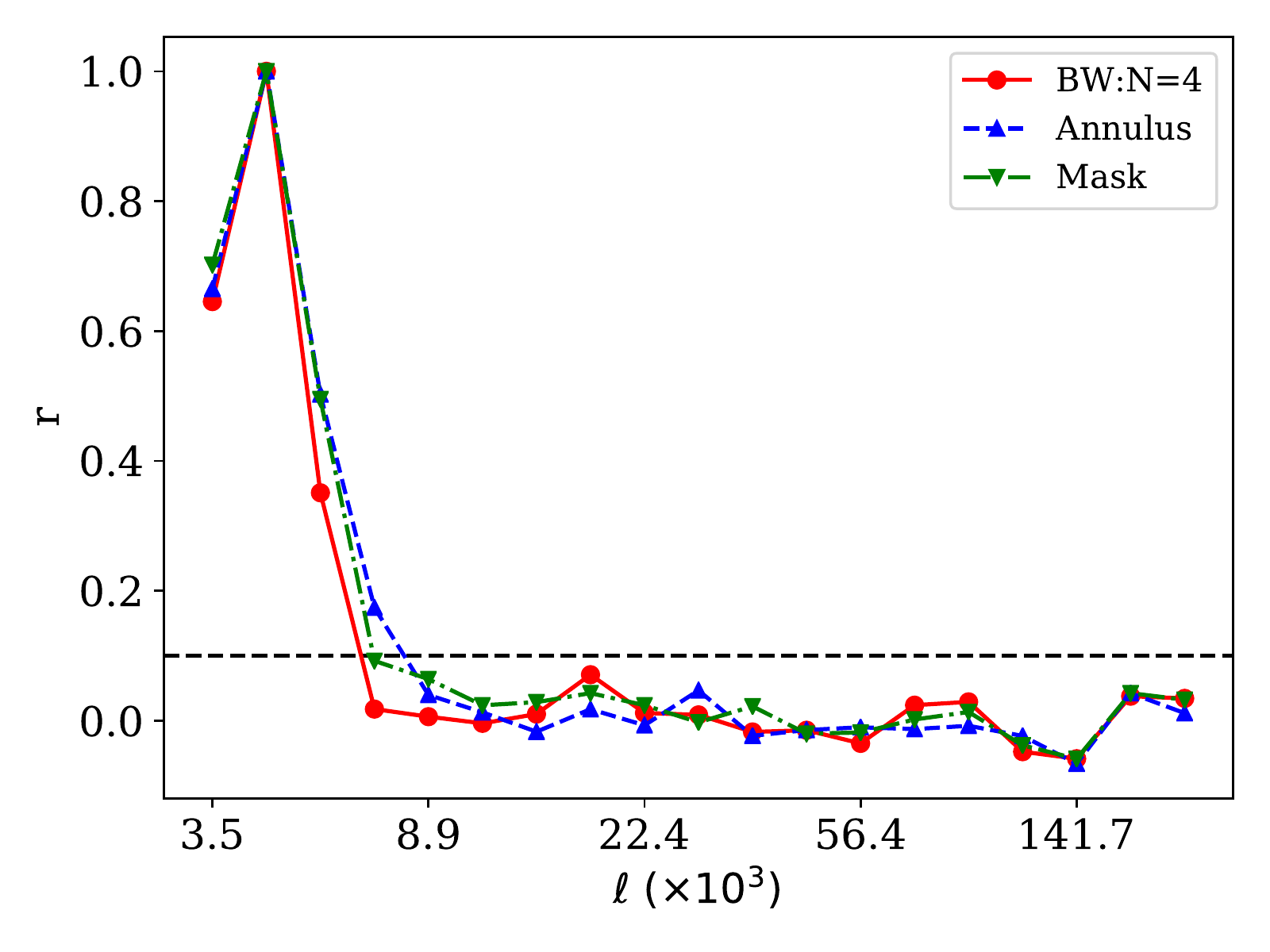}
\includegraphics[width=85mm,height=70mm,angle=0]{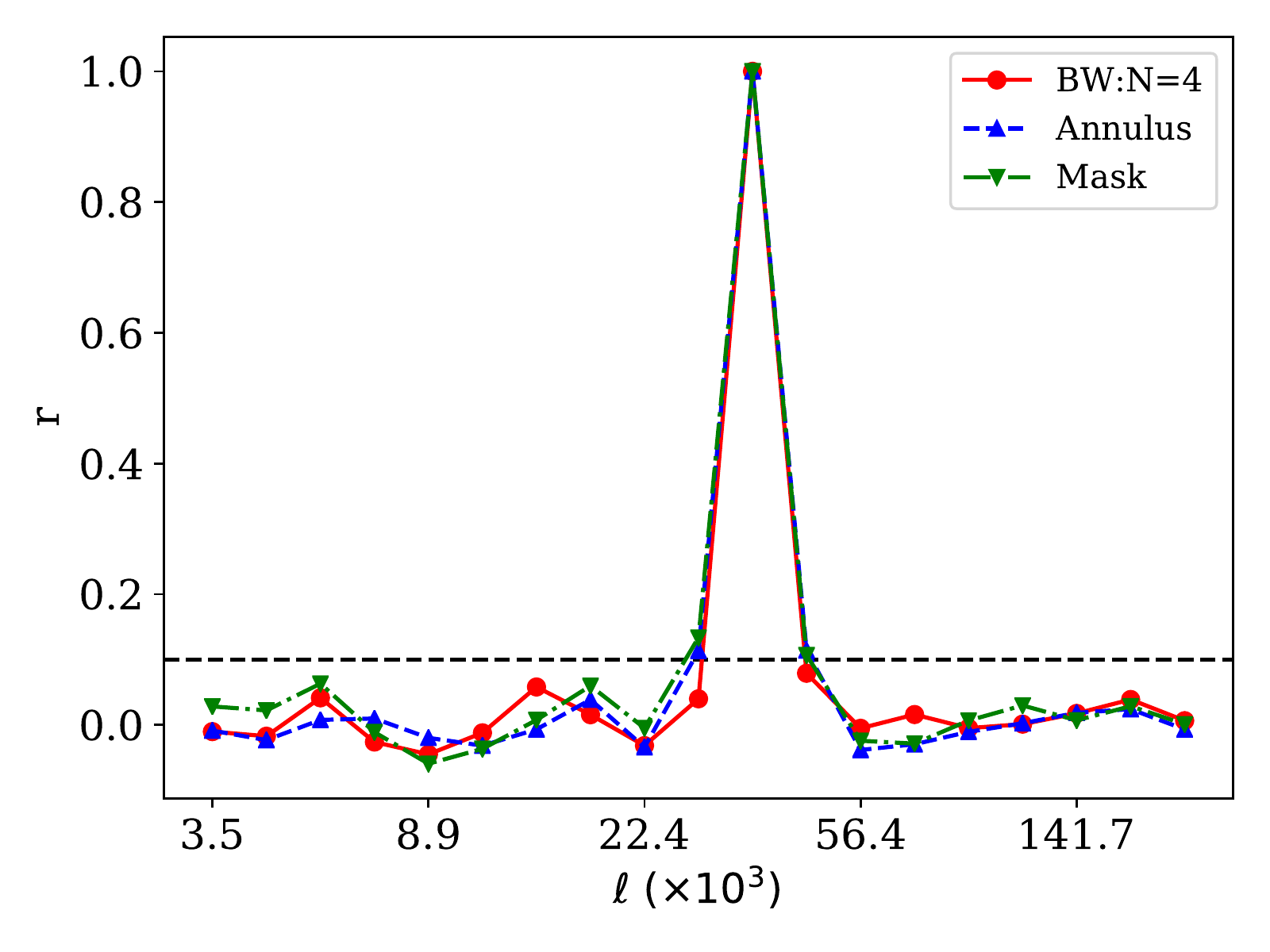}
\caption{This shows the correlation coefficient $(r_{ij})$ for three
  different window functions as indicated in the figure. The left and
  right panels respectively show the correlation between the bins
  centered at $\ell=6\times10^3$ and $5\times10^4$ with all the other
  $\ell$ bins. The horizontal dashed line shows the value
  $r_{ij}=0.1$.}
\label{fig:fig3a}
\end{center}
\end{figure}

The Mask and the Annulus window functions have smaller angular
  features as compared to the BW window and we expect a larger
  correlation between different $\ell$-bins. Here also we have studied
  these correlations using $1,000$ independent realizations of the
  simulations.  The left and right panels of Figure \ref {fig:fig3a}
  respectively show the correlation coefficients $(r_{ij})$ for the
  bins centered at $\ell=6\times10^3$ and $\ell=5\times10^4$ with all
  the other $\ell$-bins.  In addition to the Mask and Annulus windows,
  for comparison, the results are shown for the BW window with
  $N=4$. For all three windows, in the left panel $(\ell < 9 \times
  10^3)$ we find that the two adjacent bins show correlations. For the
  Annulus window, one further bin $(\ell=7.1 \times 10^3)$ shows a
  correlation of around $r_{ij}\sim 0.2$, the correlation coefficient
  is small $\mid r_{ij} \mid \le 0.1$ for all the other bins in the
  left panel. Further, in the right panel the correlation with all the
  other bins has values $\mid r_{ij} \mid \le 0.1$.  We see that the
  Annulus and Mask window functions do introduce some correlations.
  Like the Gaussian and BW window with $N=4$, the effect is restricted
  to adjacent $\ell$-bins and low $\ell$ values $(\ell \le 9 \times
  10^3)$.

\section{NGC~628 Data}
\label{data}
\begin{figure}
\begin{center}
\includegraphics[width=55mm,angle=0]{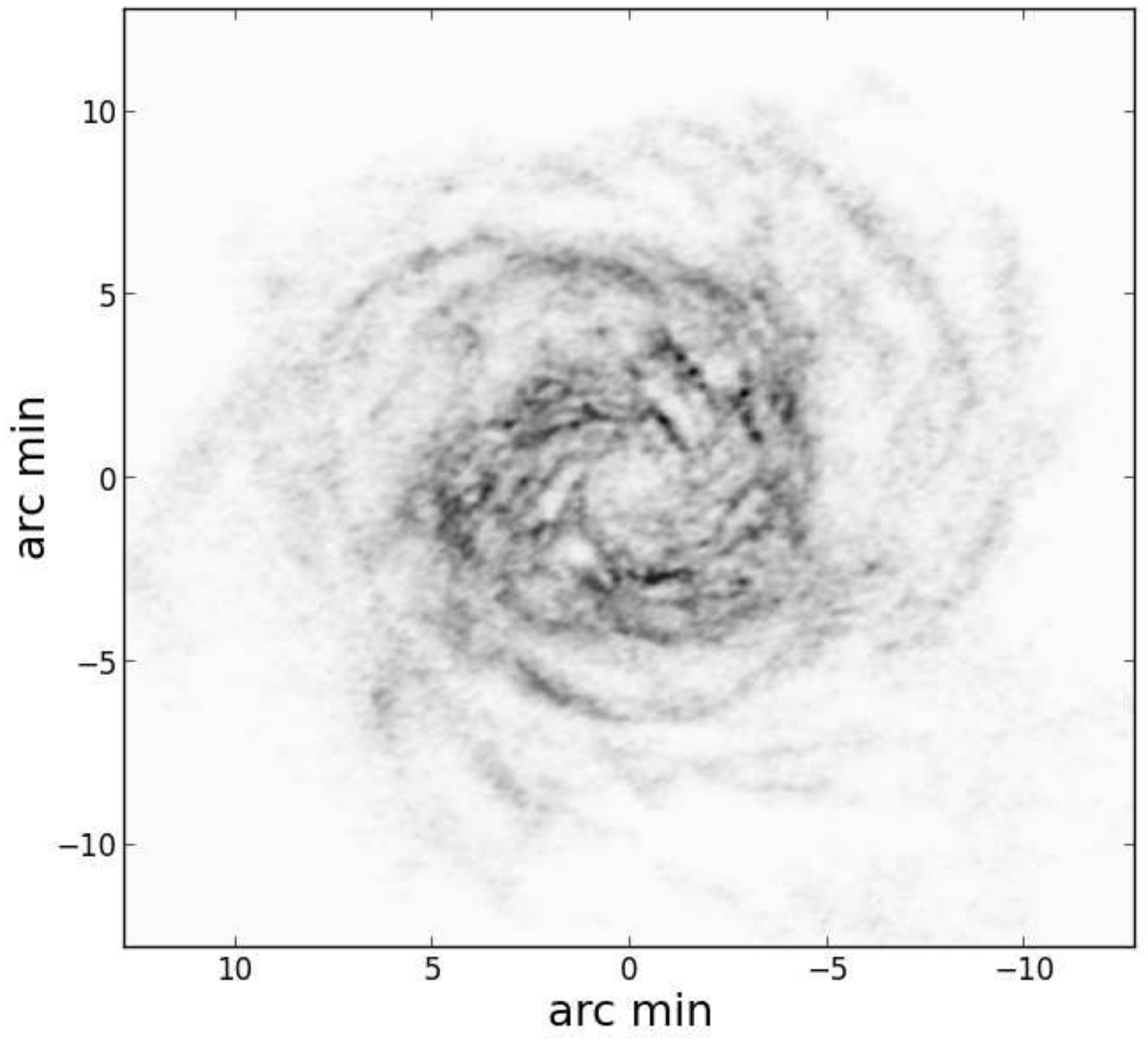}
\caption{This shows the moment0 \HI~map of the galaxy NGC~628 using
  data from the THINGS survey.}
\label{fig:fig4}
\end{center}
\end{figure}

\citet{walter08} have observed the \HI~emission from 34 spiral
galaxies using B, C and D array configurations of the Very Large Array
(VLA) as a part of The HI Nearby Galaxy Survey (THINGS). For our
analysis, we use the interferometric \HI~data of the galaxy NGC~628
from THINGS. NGC~628 is an almost face-on spiral galaxy with an
average inclination angle of $15^{\circ}$ at a distance of $7.3$ Mpc
\citep{deblok08}. The \HI~extent of the galaxy is $22.0^{'} \times
20.0^{'}$. We use the interferometric data with the primary
calibration from the THINGS archive and use the Astronomical Image
Processing System (AIPS) for further analysis. We model the
synchrotron continuum of the galaxy using visibilities from the
channels without \HI~emission and make a continuum image. We perform a
few rounds of self-calibration to improve the signal to noise ratio in
the continuum image, and then do a continuum subtraction using the
AIPS task UVSUB to retain only the \HI~emission in the visibilities
for further analysis. Figure \ref {fig:fig4} shows the moment0 \HI~map
of the galaxy NGC~628.

\begin{figure}
\begin{center}
\includegraphics[width=55mm,angle=0]{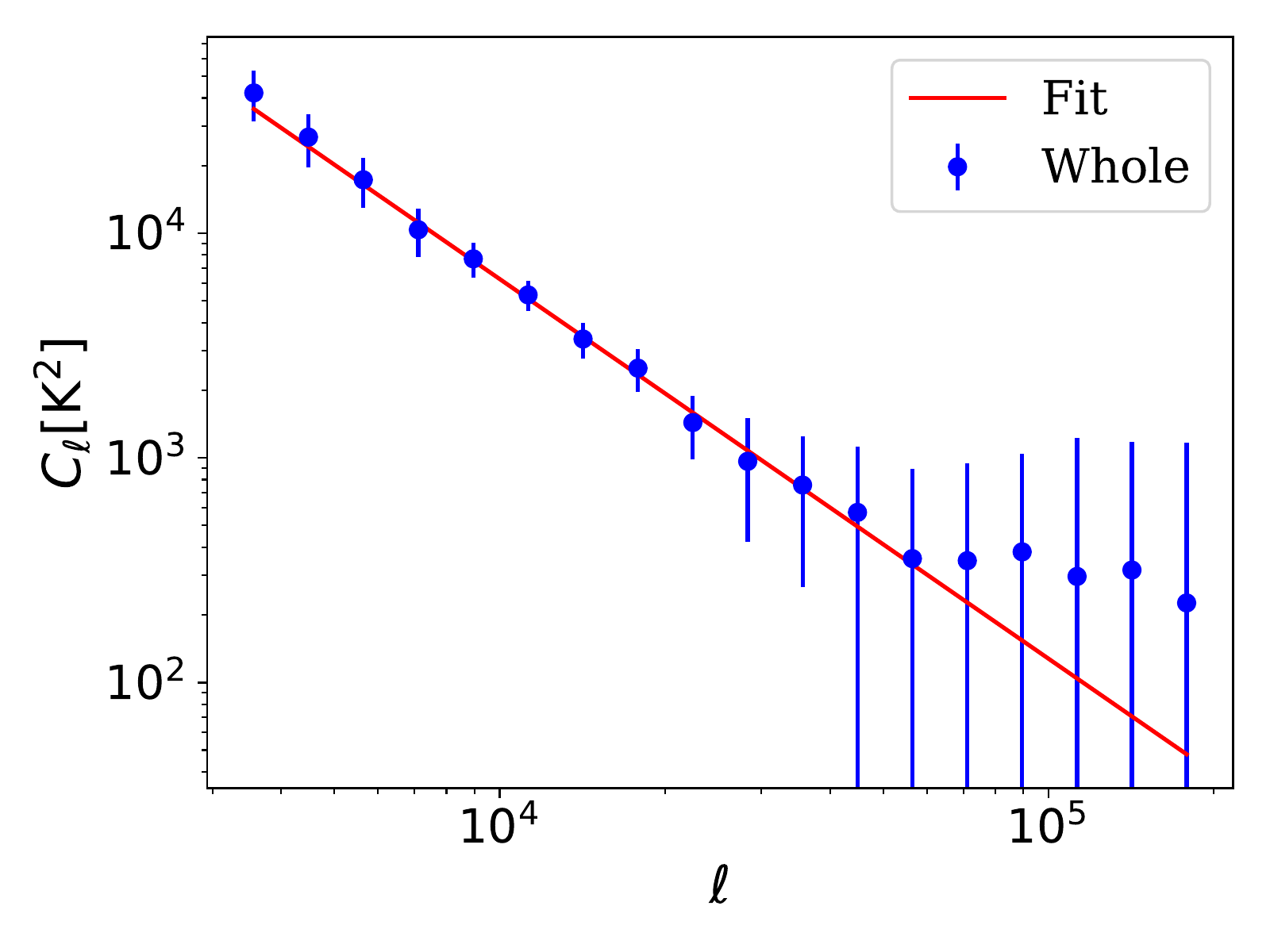}
\includegraphics[width=55mm,angle=0]{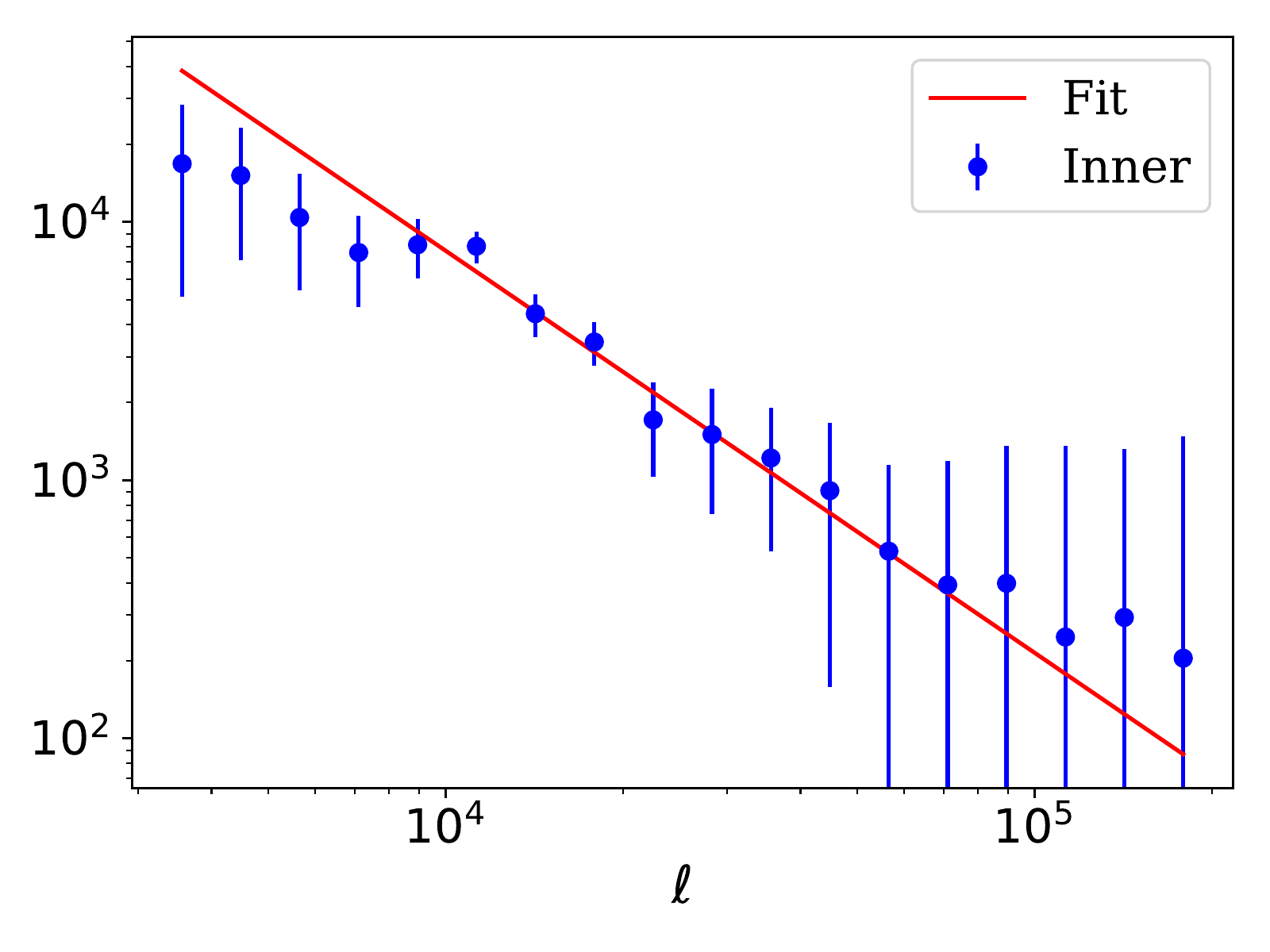}
\includegraphics[width=55mm,angle=0]{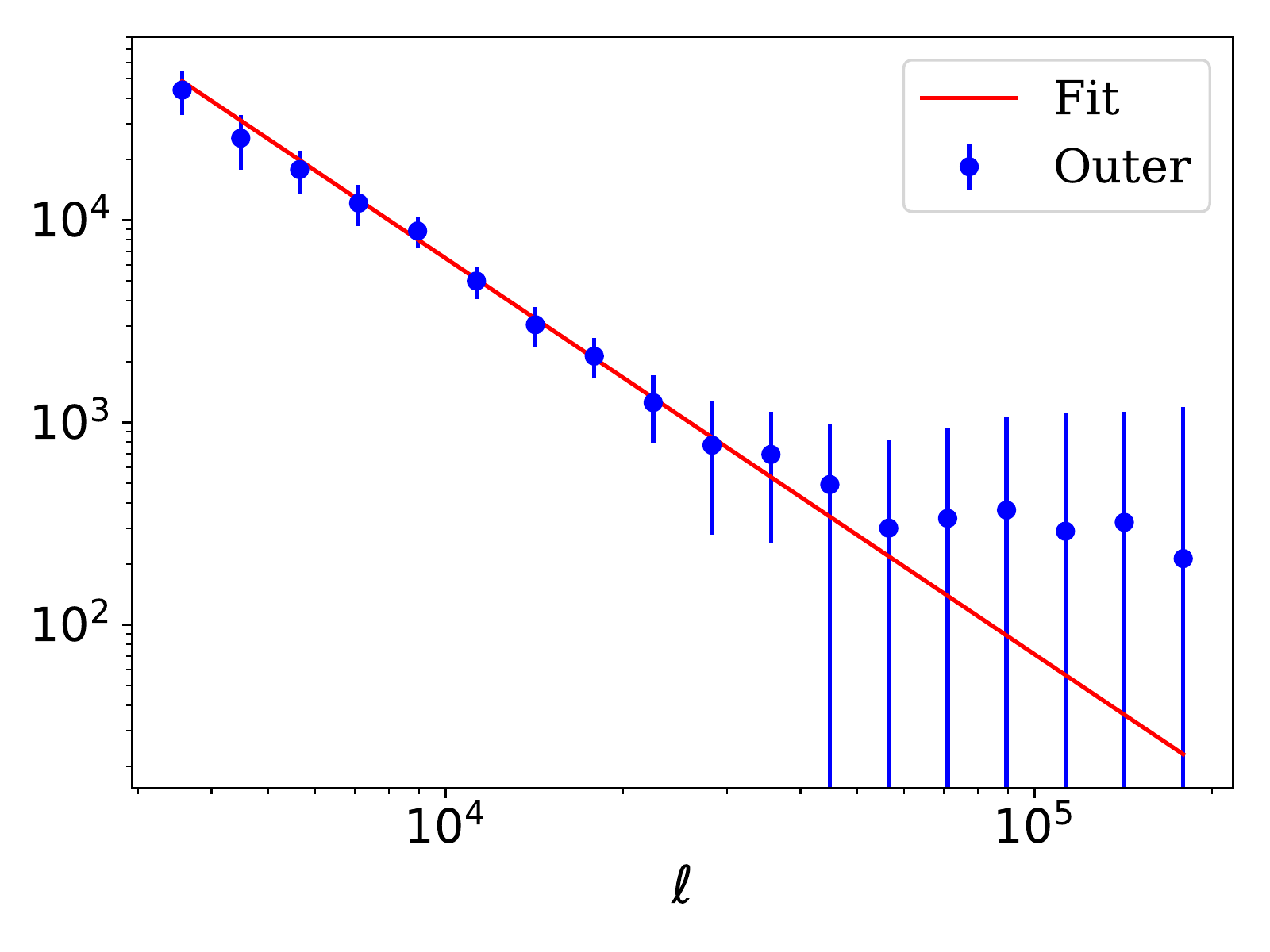}
\caption{This shows the measured $C_{\ell}$ with $1-\sigma$ error bars
  for the galaxy NGC~628. The left panel is for a BW window with
  $\theta_b=7.5^{'}$ which approximately corresponds to the entire
  extent of the galaxy.  Here, the blue points are the measured values
  and the red solid line shows the best fit model in the range
  $6\times10^3\le\ell\le6\times10^4$ (for middle panel
  $10^4\le\ell\le6\times10^4$). The middle panel shows the same but
  for a different value of $\theta_b=3.5^{'}$ which corresponds to the
  inner region of the galaxy.  The right panel is for the Annulus
  window with inner HWHM radius $3.5^{'}$ and outer HWHM radius
  $7.5^{'}$ (window shown in the left panel of Figure \ref {fig:fig3})
  which corresponds to the outer region of the galaxy. In all three
  cases, we use the value of $N=4$.}
\label{fig:fig5}
\end{center}
\end{figure}

We apply ITGE to estimate $C_{\ell}$ for the galaxy NGC~628. Figure
\ref {fig:fig5} shows the measured $C_{\ell}$ with $1-\sigma$ error
bars for the different windows mentioned in the previous section. The
left panel shows the results for a BW window with
$\theta_b=\theta_{HWHM}=7.5^{'}$ and $N=4$. This selects nearly the
entire region of the galaxy whose \HI~extent is approximately
$20^{'}$. The measured $C_{\ell}$ and the r.m.s. fluctuations $\sigma$
are shown with blue points. We have used simulations (as discussed in
Section~\ref{valid}) to estimate $\sigma$ which is the
r.m.s. fluctuation of the estimated $C_{\ell}$. These simulations
differ from those in Section~\ref{valid} in that the \HI~signal does
not fill the entire sky but is restricted to a finite angular extent
corresponding to that of the galaxy. We have incorporated this by
multiplying the sky image with a window which roughly mimics the
galaxy's radial profile, the simulated visibilities were calculated
using the resulting image.  We also add a Gaussian random noise with
standard deviation $\sigma_n=1.03 \, {\rm Jy}$ to each visibility in
order to account for the system noise. The value of $\sigma_n$ is
calculated using the rms.  of the actual measured VLA visibilities
 considering the large baselines which are likely
to be noise dominated, and we have used $128$ independent
realizations of the simulations to estimate $\sigma$.

Considering the estimated $C_{\ell}$, we identify a region in
$\ell$-space where $C_{\ell}$ is likely to be dominated by the
galaxy's \HI~signal and fit a power law (eqn. \ref{eq:b1}) to the
measured values. The $\ell$ range we used for fitting is
$6\times10^3\le\ell\le6\times10^4$.  At smaller $\ell$ values the
measured $C_{\ell}$ is affected by the finite angular extent of the
galaxy and the window function whereas the system noise becomes large
at the larger $\ell$ values beyond this range.  The best fit values
for $A$ and $\beta$ are $(3\pm0.3)\times10^5$ and $1.7\pm0.04$.
  We have used values of $A$ and $\beta$ close to these best-fitted
  values to simulate the sky for the validation of ITGE in
  Section~\ref{valid}. The best fit model is shown with a red solid
line in the left panel of Figure \ref {fig:fig5}. The measured $\beta$
is roughly consistent with the earlier measurement by
\citet{dutta13na} where they obtained $\beta=1.6\pm0.1$. As discussed
in \citet{dutta13na}, we can interpret this power-law nature of the
$C_{\ell}$ is due to the two-dimensional ISM turbulence in the plane
of galaxy's disk.

The middle panel of Figure \ref {fig:fig5} shows the estimated
$C_{\ell}$ for a smaller BW window of width $\theta_b=3.5^{'}$ and we
use the same value of $N=4$ as earlier. Here, the window blocks out
the outer parts of the galaxy, and we concentrate only the central
part of the galaxy. As the width of the window is smaller here,
  the effect of the convolution extends upto $\ell\sim10^4$. In this
  case we use the $\ell$-range $10^4\le\ell\le6\times10^4$ to fit the
  measured $C_{\ell}$. The best fit values of the parameters are
  $A=(3\pm1)\times10^5$ and $\beta=1.55\pm0.1$.

The right panel of Figure \ref {fig:fig5} shows the estimated
$C_{\ell}$ for the Annulus window with inner HWHM radius $3.5^{'}$ and
outer HWHM radius $7.5^{'}$. Here our aim is to measure the \HI~signal
from only the outer region of the galaxy. In this case the best
  fit value of the parameters are $A=(6\pm0.9)\times10^5$ and
  $\beta=2.0\pm0.06$. Here we see a difference in the power law index
  between the inner part and the outer parts of the galaxy.  This
signifies that the statistical properties of the \HI~fluctuations are
different in the central part of the galaxy where star formation is
taking place as compared to the outer parts. We plan to investigate
this issue in more detail using higher sensitivity data of the same
galaxy as well as repeating a similar analysis for other galaxies.

\section{Summary and conclusions}
\label{summ}
It is often useful to taper the sky response when estimating the power
spectrum of the diffuse sky signal directly from the visibilities
measured in radio interferometric observations. In some contexts, it
may also be desirable to mask out the sky signal from certain
directions in order to restrict the analysis to select regions of the
sky. For example, we may wish to avoid certain regions of the sky
which are contaminated by strong foreground residuals, or we may be
interested in studying the power spectrum of the signal from just the
outer parts of a galaxy. In this paper, we introduce the Imaged based
Tapered Gridded Estimator (ITGE) for estimating the power spectrum
directly from the measured visibilities.  Here it is possible to
modulate the sky response through a window function which is
implemented in the image domain on the sky plane, and it is possible
to implement a wide variety of window functions in contrast to an
earlier version which was purely visibility based
\citep{samir16b}. The ITGE deals with gridded data (both visibility
and image) and therefore is computationally efficient. The ITGE has an
added feature that it internally estimates the system noise
contribution and exactly subtracts this out thereby providing an
unbiased estimate of the power spectrum.

We have validated the ITGE using realistic $1.4{\rm GHz}$ simulations
of VLA observations at a single frequency considering a power law
input model angular power spectrum.  We have considered a variety of
window functions (e.g. the Gaussian, Butterworth (BW), Annulus and
Mask windows) and we show that we are able to recover the input model
angular power spectrum quite accurately for all of the window
functions that we have considered except one BW window with $N=256$
for which the window function falls very sharply beyond the HWHM
(Figure \ref {fig:fig1}). However, the BW window with $N=2,4$ where
there is a more gradual decline beyond the HWHM is found to work quite
well. We conclude that the ITGE is able to faithfully quantify the
angular power spectrum over a $\ell$ range which depends on the
angular extent of the window provided that the window function does
not have any very sharp features. We have also studied the error
  covariace between different $\ell$ bins and found that there is some
  correlation $(r_{ij}\sim 0.5)$ between the adjacent bins at small
  $\ell$ $( \le9\times10^3)$, whereas the $\ell$-bins are all
  uncorrelated at larger $\ell$ where $\mid r_{ij} \mid$ has values
  $\sim 0.1$ or smaller.

We have applied the ITGE to estimate the angular power spectrum
$C_{\ell}$ of the \HI~21-cm emission from the galaxy NGC~628. We have
carried out the analysis using three different windows; the first two
restrict the sky response to discs of HWHM radius $7.5^{'}$ and
$3.5^{'}$ which respectively correspond to the entire extent and the
inner region of the galaxy. The third window restricts the sky
response to the annulus bounded by the two discs mentioned above, and
this essentially quantifies the power spectrum of the \HI~emission
from the outer parts of the galaxy.  For all the cases we identify the
$\ell$ range which is likely to be dominated by the galactic \HI~and
where we expect ITGE to faithfully quantify the angular power
spectrum, and we fit a power law to the $C_{\ell}$ measured in this
range. We find that the best fit values of the power law index has
values $\beta=1.7 \pm 0.04$ and $\beta=1.55 \pm 0.1$ for the entire
galaxy and the inner region respectively, both of which are consistent
with an earlier measurement of the angular power spectrum of the
entire galaxy \citep{dutta13na} who find $\beta=1.6 \pm 0.1$ and
interpret this as arising from two-dimensional turbulence in the plane
of the galaxy.  The slope in the outer region of the galaxy is however
found to be $\beta=2.0 \pm 0.06$ which is significantly different.
This signifies that the statistical properties of the \HI~fluctuations
in the outer region of the galaxy are possibly different from those at
the central part of the galaxy where star formation is taking place.
We plan to investigate this issue using higher sensitivity data as
well carry out a similar analysis for other galaxies in future.

Further, the entire analysis here is restricted to a single
frequency. It is quite straight-forward to extend this to
multi-frequency data through the multi-frequency angular power
spectrum $C_{\ell}(\Delta\nu)$ (MAPS:\citet{datta07}) or equivalently
the 3D power spectrum $P(\k)$ (e.g. \citealt{samir16b} ). We plan to
address these issues in future work.

\section{Acknowledgements}
We thank an anonymous referee for helpful comments. SC acknowledge
NCRA-TIFR for providing financial support. We acknowledge the THINGS
collaboration for providing the data for galaxy NGC~628.

\end{document}